\documentclass[aps,12pt,superscriptaddress,amsfonts,amssymb,amsmath]{revtex4}
\pdfoutput=1

\usepackage{graphicx}
\usepackage{epsfig}
\usepackage{makeidx}
\usepackage{epstopdf}
\usepackage{natbib}

\begin{document}

\title{High-frequency electromagnetic emission \\ from non-local wavefunctions
}

\author{G.\ Modanese \footnote{Email address: giovanni.modanese@unibz.it}}
\affiliation{Free University of Bolzano-Bozen \\ Faculty of Science and Technology \\ I-39100 Bolzano, Italy}

\linespread{0.9}

\begin{abstract}

In systems with non-local potentials or other kinds of non-locality, the Landauer-B\"uttiker formula of quantum transport leads to replace the usual gauge-invariant current density $\textbf{J}$ with a current $\textbf{J}^{ext}$ which has a non-local part and coincides with the current of the extended Aharonov-Bohm electrodynamics. It follows that the electromagnetic field generated by this current can have some peculiar properties, and in particular the electric field of an oscillating dipole can have a long-range longitudinal component. The calculation is complex because it requires the evaluation of double-retarded integrals. We report the outcome of some numerical integrations with specific parameters for the source: dipole length $\sim 10^{-7}$ cm, frequency 10 GHz. The resulting longitudinal field $E_L$ turns out to be of the order of $10^2$ to $10^3$ times larger than the transverse component (only for the non-local part of the current). Possible applications concern the radiation field generated by Josephson tunnelling in thick SNS junctions in YBCO and by current flow in molecular nano-devices.

\end{abstract}

\maketitle

\section{Introduction}
\label{introduction}

The extended Maxwell equations by Aharonov and Bohm (\cite{ohmura1956new,aharonov1963further,van2001generalisation,jimenez2011cosmological,hively2012toward,Modanese2017MPLB,modanese2017electromagnetic,arbab2017extended,modanese2019design,hively2019classical}; see also eq.s (\ref{eqE}), (\ref{eqB}) in the Appendix) are employed for the calculation of electromagnetic fields generated by sources which violate the local charge conservation condition $\partial_t \rho+\nabla  \cdot {\bf{J}} = 0$. Barring exceptional situations in cosmology where such violations may occur at the macroscopic level, a possible microscopic failure of local conservation has been predicted in quantum mechanics in the following situations:

\begin{enumerate}
	\item In systems described by fractional quantum mechanics \cite{laskin2002fractional,Lenzi2008fractional,zhang2015propagation,wei2016comment,zhang2017unveiling,petreska2019time,modanese2018time}.
	\item In ordinary quantum mechanics, in the presence of non-local potentials \cite{baraff1998model,ferry1999complex,chamon1997nonlocal,balantekin1998green,latora1999superdiffusion,caspi2000enhanced,Lenzi2008solutions,sandev2014time,sandev2016effective,modanese2018time}, and in particular in first-principles calculations of transport properties using density functional theory and non-equilibrium Green functions \cite{li2008definition,zhang2011first,dreyer2018current}. The latter approach has been very successful for the exact description of quantum transport in nano-devices, which is otherwise not viable in terms of local quantum field theories.
	\item For the proximity effect in superconductors, especially in thick SNS junctions in cuprates, where the Gorkov equation cannot be properly approximated by a local Ginzburg-Landau equation \cite{hook1973ginzburg,hilgenkamp2002grain,modanese2018time,modanese2019design}.
\end{enumerate}

Concerning Point 2, we recall that the Landauer-B\"uttiker formula for the current in quantum transport, when  applied to wavefunctions in the presence of a non-local potential \cite{li2008definition,zhang2011first}, inevitably leads to the definition of a non-local charge density $\rho^{ext}$ and current density $\textbf{J}^{ext}$ which differ from the usual gauge-invariant expression, and coincide with those of the extended Aharonov-Bohm electrodynamics, namely 
\begin{align}
	\rho^{ext}=\rho+\rho^{non-loc}=\rho-\frac{1}{4\pi c^2}\frac{\partial}{\partial t}\int d^3y \frac{I\left(t_{ret},\textbf{y} \right)}{\left|\textbf{x}-\textbf{y} \right|} \label{rho-ext}
\end{align}
\begin{align}
	\textbf{J}^{ext}=\textbf{J}+\textbf{J}^{non-loc}=\textbf{J}+\frac{1}{4\pi c} \nabla \int d^3y \frac{I\left(t_{ret},\textbf{y} \right)}{\left|\textbf{x}-\textbf{y} \right|} \label{j-ext}
\end{align}
where $t_{ret}=t-c^{-1}|\textbf{x}-\textbf{y}|$ and the ``extra-source'' $I(t,\textbf{x})$ is the function which quantifies the violation of local current conservation:
\begin{align}
	I(t,\textbf{x})=\frac{\partial\rho}{\partial t}+\nabla\cdot\textbf{J}
\end{align}
\begin{align}
	\rho=|\Psi|^2; \qquad \textbf{J}=\frac{{ - i\hbar }}{{2m}}\left( {{\Psi ^*}\nabla \Psi  - \Psi \nabla {\Psi ^*}} \right)
\end{align}

The current $\textbf{J}$, which can be interpreted as $\sim \rho \textbf{v}$ in a classical limit, is locally non-conserved and has in this case ``sources and sinks'' which are, however, invisible to an electromagnetic probe (this is the so-called ``censorship property'' of Aharonov-Bohm electrodynamics and constitutes a safeguard of the locality of the electromagnetic field).

Other authors (\cite{dreyer2018current} and refs.) define the extended current in a different way from Refs.\ \cite{li2008definition,zhang2011first}, and take into account the possibility of adding to it a solenoidal component. The correct definition of the physical current is still an open question, also regarding the dissipation properties of the non-local part: should the latter be interpreted as a ``virtual'' current or as a real current with real dissipation? In this context, a detailed calculation and experimental verification of the predictions of Aharonov-Bohm extended electrodynamics would clearly be of special interest.

In this work we are concerned with the computation of the electromagnetic field generated by the non-local part of the current. This field is independent from any solenoidal component, and therefore the ambiguities mentioned above do not directly affect our results. It turns out that the radiation field generated by an oscillating dipole with a failure in local conservation (the most obvious example, apart from the quasi-static case examined in \cite{modanese2019design}) has very interesting features: namely, it contains an anomalous longitudinal electrical component, with large strength and long range. 

For the frequency considered (10 GHz) we found that the strength of the longitudinal component at a distance between 3$\lambda$ and 13$\lambda$ is of the order of $10^2$ to $10^3$ times the standard transverse component. This factor must be weighted with a small factor that measures the importance of the non-local current in comparison to the standard current. According to \cite{li2008definition}, first principles calculations of conventional current density can give errors for current as large as 20\% for molecular devices.
However, most molecular devices do not carry currents large enough to generate macroscopic fields. An exception could be graphene \cite{botello2011quantum}. Other materials which exhibit macroscopic quantization, large currents and possibly non-local currents are, as mentioned, cuprate superconductors. 

The computation of the radiation field is technically very difficult due to the presence of double-retarded integrals and ``secondary sources'' $\rho^{ext}$, $\textbf{J}^{ext}$ extended in space. So we had to resort to a complex integro-dipolar expansion and to long 6-dimensional Monte Carlo integrations, obtaining numerical results for some fixed values of the source parameters, chosen in view of plausible experimental situations.

It is likely that in future developments the finite-elements integration techniques currently used for the standard Maxwell equations can be extended to Aharonov-Bohm electrodynamics, but this extension is far from obvious, because the familiar vector-analysis features of the Maxwell equations are strongly affected by the removal of the local charge conservation condition. Therefore any technique based on the usual properties of the divergence of $\textbf{E}$ and circuitation of $\textbf{B}$ must be reconsidered, and in a first approach we deemed it safer to use only the retarded integral solutions, which for the non-local part of the sources can be written in terms of the potentials as (we set $k=c^{-1}$):
\begin{align}
	\phi^{non-loc}=\frac{1}{4 \pi} \int \frac{d^3y}{|\textbf{x}-\textbf{y}|} \left[ -k^2 \frac{\partial}{\partial t} \int \frac{d^3z}{|\textbf{y}-\textbf{z}|} I\left(t - k|\textbf{y}-\textbf{z}|,\textbf{z} \right) \right]_{t \to t-k|\textbf{x}-\textbf{y}|}
	\label{phiI}
\end{align}
\begin{align}
	\textbf{A}^{non-loc}=\frac{1}{4 \pi} \int \frac{d^3y}{|\textbf{x}-\textbf{y}|} \left[ k \nabla_y \int \frac{d^3z}{|\textbf{y}-\textbf{z}|} I\left(t - k|\textbf{y}-\textbf{z}|,\textbf{z} \right) \right]_{t \to t-k|\textbf{x}-\textbf{y}|}
	\label{AI}
\end{align}
In the following the suffix \emph{non-loc} will be omitted.

The extra-source $I(t,{\bf x})$ is represented by two opposite Gaussian peaks which can have spherical or ellipsoidal symmetry. This choice is based on Ref.\ \cite{modanese2018time}, where we have found $I$ explicitly from the solutions of fractional wave equations and of wave equations with non-local potential.

The paper is organized as follows. In Sect.\ \ref{sec2} we first recall a formal argument showing that the extended equations in vacuum can have solutions with a longitudinal propagating component; then we define the non-conserved dipolar source used for the numerical calculation, we list the formal steps necessary for computing the electric field and we illustrate the method followed in the Monte Carlo integration. In Sect.\ \ref{6d} we set out a new integro-dipolar expansion which is needed in order to eliminate from the numerical integrations the large opposite fluctuations due to the monopolar terms. In Sect.\ \ref{6d} we compute the electric field generated by a conserved source which serves as a benchmark for the amplitude of the anomalous longitudinal component. Sects.\ \ref{res} and \ref{conc} contain our results and conclusions.

\section{Oscillating dipolar source and integral expressions for the radiation field}
\label{sec2}

In most papers on extended Maxwell equations it is noticed that, unlike the standard Maxwell equations, they admit wave solutions with a longitudinal electric component. Some authors cite experimental evidence reportedly showing the existence of electromagnetic waves with non-transverse components \cite{giakos1993detection,monstein2002observation,monstein2004remarks}. Such evidence is scarce, compared to the immense body of precision measurements and technological applications of transverse electromagnetic waves. This implies, however, that the potential practical interest for such propagation modes is large, in case their existence is confirmed. It is immediate to see how the prediction of longitudinal electromagnetic waves emerges from the extended Maxwell equations. The first Maxwell equation in vacuum states that $\nabla \cdot {\bf E}=0$, so for a plane wave ${\bf E}={\bf E}_0 e^{i({\bf k}{\bf x}-\omega t)}$ (or locally) one obtains the transversality condition ${\bf E}_0\cdot {\bf k}=0$, where ${\bf k}$ defines the propagation direction of the wave. The first equation of the extended Aharonov-Bohm theory in vacuum is instead
\begin{equation}
\nabla \cdot {\bf E}=-\frac{1}{c} \frac{\partial S}{\partial t}
\end{equation}
where $S$ is a scalar field which satisfies the equation
\begin{equation}
\frac{1}{c^2}\frac{\partial^2 S}{\partial t^2}-\nabla^2 S = I = \partial_t \rho + \nabla \cdot {\bf J}
\label{boxS}
\end{equation}
The ``extra-current'' $I$ is non zero at the points where the local conservation of charge fails. If charge is locally conserved everywhere, then the $S$ field is completely decoupled from matter. In this case, even in the extended theory no longitudinal components should be expected.

Eq.\ (\ref{boxS}) can be solved for $S$, obtaining the first extended Maxwell equation in vacuum with a non-local source term:
\begin{equation}
\nabla \cdot {\bf E}=-\frac{1}{c} \frac{\partial }{\partial t} \int d^3y \frac{I\left(t_{ret},\textbf{y} \right)}{\left|\textbf{x}-\textbf{y} \right|}
\end{equation}
This shows that the divergence of ${\bf E}$ in vacuum is equal to a term that we can call ``secondary charge density'' or ``cloud charge'', generated in the surrounding space by the local non-conservation of the ``primary current''. Therefore in a wave solution in vacuum the electric field can have a longitudinal component.

In order to find the concrete predictions of the theory and assess the feasibility of an experimental check, it is necessary to compute exactly the longitudinal electric radiation field $E_L$ generated by an appropriate source, compare its magnitude order with that of the transverse field $E_T$ and make sure that it does not vanish for some reason not apparent from the general form of the equations. Symmetry can play a crucial role here. We have previously proven in \cite{modanese2019design}, for instance, that in the case of a quasi-stationary extra-source $I$ representing a Josephson weak link with local non-conservation, the anomalous magnetic field generated by $I$ is zero, and there is indeed an observable effect because the corresponding Biot-Savart field is missing. This happens, however, for a source $I$ with spherical symmetry; otherwise the anomalous field partially replaces the missing Biot-Savart field.

\subsection{Steps needed to write the integral expression for the electric field
}

\begin{figure}
\begin{center}
\includegraphics[width=11cm,height=8cm]{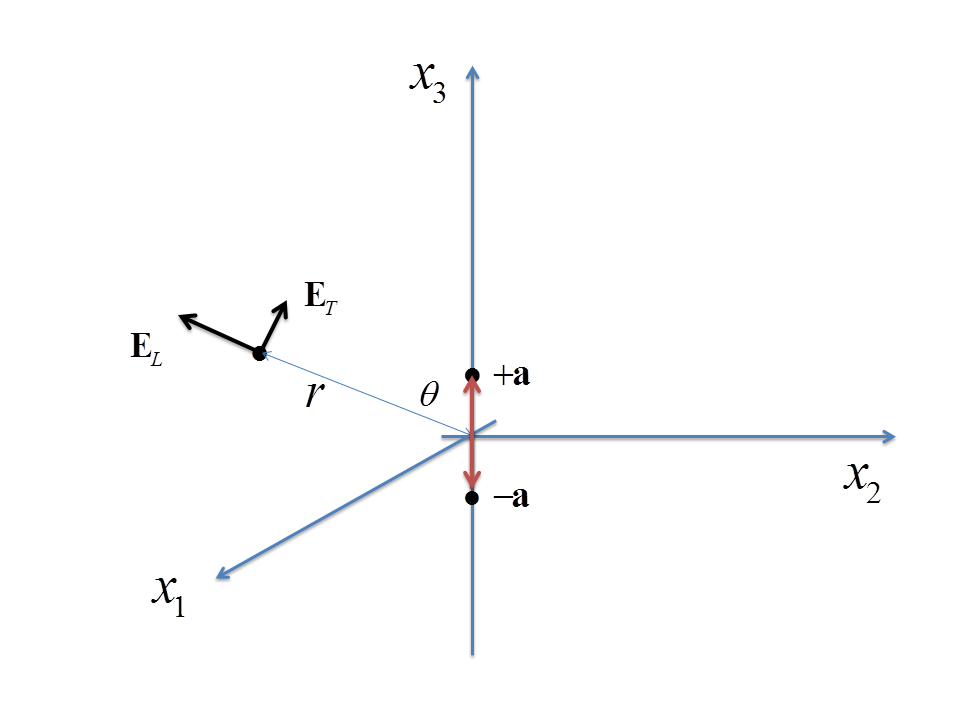}
\caption{Geometrical setting for the calculation of the extra-current $I$ (\ref{extra-cur}) produced by the failure of local conservation. The charge $q$ oscillates between points placed on the $x_3$-axis, at $-{\bf a}$ and $+{\bf a}$. The field is computed in the plane $x_1$--$x_3$, with angle $\theta=45^{\circ}$.} 
\label{fig-assi}
\end{center}  
\end{figure}

With reference to Fig.\ \ref{fig-assi}, consider an oscillating dipolar source with two opposite charges at ${\bf x}=-{\bf a}$ and ${\bf x}=+{\bf a}$, of the following form:
\begin{equation}
\rho(t,{\bf x} )=q\cos(\omega t) f({\bf x} ); \ \ \ {\bf J}=0
\label{dd-source}
\end{equation}
where $f({\bf x} )$  is essentially a regularized double-$\delta$, whose support can be adapted to describe a sphere or a disk (see below, eq.\ (\ref{doppia-exp}))
\begin{equation}
f({\bf x} )\simeq \delta^3( {\bf x}-{\bf a} )-\delta^3( {\bf x}+{\bf a} )
\end{equation}
The absence of current (${\bf J}=0$) violates local conservation and can be described as the consequence of a ``strong-tunnelling'' process \cite{modanese2018time}. In a real source, only a small part of the total charge will oscillate without a current, so we are focussing our attention on the field generated by that part.

In order to compute the field of the source (\ref{dd-source}) using the extended Maxwell equations we must write the potentials $\phi$ and ${\bf A}$ as double-retarded integrals like in eqs.\ (\ref{phiI}), (\ref{AI}), and then we have ${\bf E}={\bf E}^\phi+{\bf E}^A=-\nabla \phi-k \partial_t {\bf A}$.

The integrand in eqs.\ (\ref{phiI}), (\ref{AI}) is given by $I=\partial_t \rho+\nabla \cdot {\bf J}$; therefore, since ${\bf J}=0$, one has here
\begin{equation}
I(t,{\bf x} )=\partial_t \rho(t, {\bf x} )=-q\omega \sin(\omega t) f( {\bf x} )
\label{extra-cur}
\end{equation}

The steps needed to obtain the contribution $E_i^\phi$ are then the following:
\begin{itemize}
\item Retardate $t \to t-k|{\bf y}-{\bf z}|$ in $I(t,{\bf z} )$, divide $I$ by $|{\bf y}-{\bf z}|$ and integrate in $d^3z$.
\item Differentiate with respect to $t$ and multiply by $(-k^2)$.
\item Retardate $t \to t-k|{\bf x}-{\bf y}|$.
\item Multiply by $1/|{\bf x}-{\bf y}|$ and integrate in $d^3y$.
\item Differentiate with respect to $x_i$ and multiply by $(-1)$.
\end{itemize}

The steps needed to obtain the contribution $E_i^A$ are the following:
\begin{itemize}
\item Retardate $t \to t-k|{\bf y}-{\bf z}|$ in $I(t,{\bf z} )$, divide $I$ by $|{\bf y}-{\bf z}|$ and integrate in $d^3z$.
\item Differentiate with respect to $y_i$ and multiply by $(-k)$.
\item Retardate $t \to t-k|{\bf x}-{\bf y}|$.
\item Multiply by $1/|{\bf x}-{\bf y}|$ and integrate in $d^3y$.
\item Differentiate with respect to $t$ and multiply by $(-k)$.
\end{itemize}

Through these steps one arrives, after long but straightforward manipulations, at the following expression for the electric field, as a double retarded integral:
\begin{equation}
E_i(t,{\bf x})=E_i^\phi(t,{\bf x}) + E_i^A(t,{\bf x})
\end{equation}
where $E_i^\phi(t,{\bf x})$ is the contribution of the scalar potential:
\begin{equation}
E_i^\phi(t,{\bf x})=-qK^2 \int d^3z \int d^3y \frac{f({\bf z})(x_i-y_i)}{|{\bf y}-{\bf z}||{\bf x}-{\bf y}|^3} \left(K|{\bf x}-{\bf y}|\sin\Omega -\cos\Omega \right)
\label{eq14}
\end{equation}
and the contribution of the vector potential is
\begin{equation}
E_i^A(t,{\bf x})=-qK^2 \int d^3z \int d^3y \frac{f({\bf z})(y_i-z_i)}{|{\bf y}-{\bf z}|^3|{\bf x}-{\bf y}|} \left(K|{\bf y}-{\bf z}|\sin\Omega -\cos\Omega \right)
\label{eq15}
\end{equation}
Here $K$ is the wavenumber: $K=k\omega=c^{-1}\omega=2\pi\lambda^{-1}$; the phase $\Omega$ is given by
\begin{equation}
\Omega=\omega (t-k|{\bf x}-{\bf y}|-k|{\bf y}-{\bf z}|)
\end{equation}
and $f({\bf z})$ is a regularized representation of the double $\delta$-function of the dipolar source in eq.\ (\ref{dd-source}) (because, as discussed in \cite{modanese2018time}, extra-sources originating from a non-local wavefunction are smooth):
\begin{equation}
f({\bf z})=\frac{1}{\sqrt{(2\pi)^3}\varepsilon d^2}\left[ e^{ -\frac{1}{2}\left( \frac{z_1^2}{d^2}+\frac{z_2^2}{d^2}+\frac{(z_3-a)^2}{\varepsilon^2} \right) } - e^{ -\frac{1}{2}\left( \frac{z_1^2}{d^2}+\frac{z_2^2}{d^2}+\frac{(z_3+a)^2}{\varepsilon^2} \right) } \right]
\label{doppia-exp}
\end{equation}
The parameter $a$ represents the length of the dipole and is taken equal to $2.5\cdot 10^{-7}$ cm (in the following, $a_1=a_2=0$; $a_3=a$). The parameter $\varepsilon$ represents the size in the 3-direction of the dipole charges, and $d$ their size in the 1- and 2-directions. At the beginning, the oscillation frequency is set to $\omega=2\pi \cdot 10^{10}$ Hz and $d=\varepsilon=10^{-7}$ cm.

The field is computed at the point ${\bf x}=\left(\frac{r}{\sqrt{2}},0,\frac{r}{\sqrt{2}} \right)$; at the beginning we set $r=10$ cm (approximately equal to three wavelengths), then $r$ is increased up to 40 cm.
The transverse and longitudinal components of the electric field at this position are defined by the expressions
\begin{equation}
E_T=\frac{1}{\sqrt{2}}(-E_1+E_3); \ \ \ E_L=\frac{1}{\sqrt{2}}(E_1+E_3)
\end{equation}

In addition to the integral in eq.\ (\ref{eq14}) there is also another contribution to $E^\phi$, due to the normal density $\rho$ (not $\rho^{non-loc}$) of the source (\ref{dd-source}). The corresponding $\phi$ is given in Sect.\ \ref{bench} (eq.\ (\ref{eq56})), in the limit when the function $f$ becomes a double delta-function. It turns out to be of the order of $E_T^{c,rms}$ and will therefore be disregarded here in the computation of $E_L$.

The presence of the Gaussian function $f(z)$ restricts the effective range of the integration in $d^3z$ approximately to $\varepsilon$ in the directions 1, 2, and to $(a+\varepsilon)$ in direction 3. Therefore in the Monte Carlo integration procedure we just set the range of ${\bf z}$ accordingly, and the corresponding integration volume is small. Setting the range of ${\bf y}$ is much more difficult, because there is no exponential cutoff in ${\bf y}$ in the integrand, but only a decrease according to a power law. So we can only proceed empirically by integrating over an increasing range $R_y$ until the result stabilizes. All our trials give a stabilization value of $R_y$ (with the parameters employed) between approx.\ 100 and 200 cm.

\section{Integro-dipolar expansion}
\label{6d}

We divide the $\textbf{y}$ integration region using cubes centered at the origin.
When we compute the contributions of the regions with $0\le y_i \le 10^{-6}$, then $10^{-6}\le y_i \le 10^{-5}$ etc.\ (values in cm), we obtain precise results up to approx.\ $10^{-5}$; then the fluctuations become large, even in long runs ($10^{11}$ to $10^{12}$ sampling points). This happens because the two opposite monopolar contributions in the integral are large, and when the sampling points are spread over bigger volumes, their cancellation is affected by large casual errors. We therefore make recourse to a dipolar expansion in the integration region far from the primary source. This is a non-standard expansion because of the presence of the double retarded integration, so it needs special care and must be cross-checked numerically by comparing its results to those of the full integral in the intermediate integration region where $|{\bf y}|$ is small enough that the fluctuations are still under control, but large enough that the assumption $|{\bf y}| \gg a$ for the dipolar expansion is valid.

Let us first consider the case of dipole charges having spherical symmetry, so that $d=\varepsilon$ in the definition of $f({\bf z})$. We rewrite the integral for $E_i^\phi$ as the sum of two integrals $E_i^{\phi,{\bf a}}$ and $E_i^{\phi,-{\bf a}}$ for the sources at ${\bf a}$ and $-{\bf a}$, in which the ${\bf z}$ variable is shifted by $-{\bf a}$ and ${\bf a}$, respectively:
\begin{equation}
E_i^{\phi}=E_i^{\phi,{\bf a}}+E_i^{\phi,-{\bf a}}
\end{equation}
For the first integral, with shift $-{\bf a}$, we define a new variable ${\bf u}={\bf z}-{\bf a}$. Define a regularized $\delta$-function for a source centered at the origin:
\begin{equation}
F({\bf u})=\frac{1}{\sqrt{(2\pi)^3}\varepsilon^3}e^{ -\frac{1}{2} \frac{{\bf u}^2}{\varepsilon^2} }
\end{equation}

The electric field generated by the scalar potential of the source at ${\bf a}$ can be written as
\begin{equation}
E_i^{\phi,{\bf a}}  =  -qK^2 \int_{M-{\bf a}} d^3u \int d^3y 
H_i^\phi({\bf x},{\bf y},{\bf u})G^\phi(t,{\bf x},{\bf y},{\bf u}+{\bf a})
\end{equation}
where
\begin{equation}
H_i^\phi({\bf x},{\bf y},{\bf u})=\frac{F({\bf u})(x_i-y_i)}{|{\bf x}-{\bf y}|^3}
\end{equation}
\begin{equation}
G^\phi(t,{\bf x},{\bf y},{\bf u}+{\bf a})=
 \frac{ K|{\bf x}-{\bf y}|\sin\Omega-\cos\Omega }{|{\bf y}-{\bf u}-{\bf a}|}
\label{A1bis}
\end{equation}
\begin{equation}
\Omega=\omega (t-k|{\bf x}-{\bf y}|-k|{\bf y}-{\bf u}-{\bf a}|)
\label{A1-2}
\end{equation}
We have symbolically denoted the integration range of ${\bf u}$ as ``$M-{\bf a}$'', meaning that it is equal to the integration range $M$ of $z$ ($-R_z \le z_i \le R_z$) shifted by a quantity $-{\bf a}$.

The function $G^\phi(t,{\bf x},{\bf y},{\bf u}+{\bf a})$ can be expanded as a term of order zero in $a=|{\bf a}|$ and a term of order 1:
\begin{equation}
G^\phi(t,{\bf x},{\bf y},{\bf u}+{\bf a})\simeq G^\phi_0(t,{\bf x},{\bf y},{\bf u})+a \cdot G^\phi_1(t,{\bf x},{\bf y},{\bf u})
\end{equation}
Actually, the small quantity in which we make the expansion is $a/|{\bf y}|$ and we therefore expect that the expansion is accurate where $|{\bf y}| \gg a$, which is what we need, as explained above.

Let us expand the factor $1/|{\bf y}-{\bf u}-{\bf a}|$ to first order in $a$. Define ${\bf v}={\bf y}-{\bf u}$. $|{\bf v}|$ is of order $|{\bf y}|$, because $F({\bf u})$ has range $\simeq \varepsilon < a$; therefore $|{\bf v}| \gg a$. In the following we denote $v=|{\bf v}|$. 

Defining 
\begin{equation}
\Delta_a=\frac{ {\bf v} \cdot {\bf a} }{v}=( {\bf y}-{\bf u} ) \cdot \frac{{\bf a} }{v}=(y_3-u_3) \frac{a}{v}
\end{equation}
we have
\begin{equation}
|{\bf v}-{\bf a}|=v\sqrt{1+\frac{a^2}{v^2}-2\frac{\Delta_a}{v} }
\end{equation}
and we find the following first order approximations:
\begin{equation}
|{\bf v}-{\bf a}|^{-1}\simeq  \frac{1}{v}\left( 1+\frac{\Delta_a}{v} \right)
\end{equation}
and
\begin{equation}
\sin\Omega\simeq \sin\Omega_0+\cos\Omega_0 \Delta \Omega
\end{equation}
where
\begin{equation}
\Omega_0=\omega(t-k|{\bf x}-{\bf y}|-k|{\bf y}-{\bf u}|)
\end{equation}
and
\begin{equation}
\Delta\Omega=-\omega k \Delta |{\bf y}-{\bf u}-{\bf a}|=-K \Delta |{\bf v}-{\bf a}|=K \Delta_a
\end{equation}
Similarly,
\begin{equation}
\cos\Omega \simeq \cos\Omega_0-\sin\Omega_0 K\Delta_a
\end{equation}

Now we can rewrite the function $G^\phi(t,{\bf x},{\bf y},{\bf u}+{\bf a} )$ as follows:
\begin{equation}
G^\phi(t,{\bf x},{\bf y},{\bf u}+{\bf a} ) = 
\frac{1}{v} \left( 1+\frac{\Delta_a}{v} \right) \left[ K|{\bf x}-{\bf y}|(\sin\Omega_0+\cos\Omega_0 K\Delta_a )-\cos\Omega_0+\sin\Omega_0 K \Delta_a \right] 
\end{equation}
Therefore in the decomposition of $G^\phi$, the part $G^\phi_0$, with the terms independent from $a$ is
\begin{equation}
G^\phi_0= \frac{1}{v} (K|{\bf x}-{\bf y}|\sin\Omega_0-\cos\Omega_0 )
\end{equation}
and the part of first order in $a$ is given by
\begin{equation}
aG^\phi_1=\frac{\Delta_a}{v} \left[ \frac{1}{v} (K|{\bf x}-{\bf y}|\sin\Omega_0-\cos\Omega_0 )+K^2|{\bf x}-{\bf y}|\cos\Omega_0+K\sin\Omega_0 \right]
\label{ex27}
\end{equation}

In the sum $E_i^{\phi}=E_i^{\phi,{\bf a}}+E_i^{\phi,-{\bf a}}$ the terms with $G^\phi_0$ cancel, because the integral over the region ``$M-{\bf a}$'' is equal to an integral over $M$, due to the short range of the function $F({\bf u})$.
The remaining term of first order in $a$ gives
\begin{equation}
E^\phi_i=-2qK^2 \int_M d^3u \int d^3y H_i^\phi({\bf x},{\bf y},{\bf u})\cdot aG^\phi_1(t,{\bf x},{\bf y},{\bf u}) + o(a^2)
\label{ex28}
\end{equation}

The electric field generated by the vector potential of the source at ${\bf a}$ can be written as
\begin{equation}
E_i^{A,{\bf a}}  =  -qK^2 \int_{M-{\bf a}} d^3u \int d^3y 
H^A({\bf x},{\bf y},{\bf u})G_i^A(t,{\bf x},{\bf y},{\bf u}+{\bf a})
\end{equation}
where
\begin{equation}
H^A({\bf x},{\bf y},{\bf u})=\frac{F({\bf u})}{|{\bf x}-{\bf y}|}
\end{equation}
\begin{equation}
G^A_i(t,{\bf x},{\bf y},{\bf u}+{\bf a})=K
 \frac{(y_i-u_i-a_i)}{|{\bf y}-{\bf u}-{\bf a}|^2}\sin\Omega
 -\frac{(y_i-u_i-a_i)}{|{\bf y}-{\bf u}-{\bf a}|^3}\cos\Omega
\label{A1}
\end{equation}
The function $G^A_i$ can be approximately decomposed in a part independent from $a$ and a part linear in $a$, as done before for $G^\phi$:
\begin{equation}
G^A_i(t,{\bf x},{\bf y},{\bf u}+{\bf a})\simeq
G^A_{0,i}(t,{\bf x},{\bf y},{\bf u})+a\cdot G^A_{1,i}(t,{\bf x},{\bf y},{\bf u})
\end{equation}
In order to find $G^A_{0,i}$ and $G^A_{1,i}$ we expand the factors present in $G^A_i$ to first order in $a$. Start with
\begin{equation}
 \frac{1}{|{\bf y}-{\bf u}-{\bf a}|^2}=\frac{1}{|{\bf v}-{\bf a}|^2}\simeq \frac{1}{v^2} \left( 1+2 \frac{\Delta_a}{v} \right)
\end{equation}
For the component $i=1$, $a_i=0$, therefore the factor $(y_i-u_i-a_i)$ does not have components of order $a$. We obtain
\begin{equation}
G^A_{i=1}\simeq K\frac{v_1}{v^2} \left( 1+2 \frac{\Delta_a}{v} \right) (\sin\Omega_0+\cos\Omega_0 K \Delta_a )
-\frac{v_1}{v^3} \left( 1+3 \frac{\Delta_a}{v} \right)
(\cos\Omega_0-\sin\Omega_0 K \Delta_a )
\end{equation}
whose first order part is
\begin{equation}
G^A_{1,i=1}=\frac{v_1}{v^2}\Delta_a \left( K\cos\Omega_0+ \frac{2}{v}\sin\Omega_0 \right)
-\frac{v_1}{v^3}\Delta_a \left( -K\sin\Omega_0+ \frac{3}{v}\cos\Omega_0 \right)
\end{equation}
The case of $i=3$ is more involved, because $a_i=a$ in that case. We write
\begin{equation}
 \frac{y_3-u_3-a_3}{|{\bf y}-{\bf u}-{\bf a}|^2}\simeq \left( \frac{v_3}{v}-\frac{a}{v} \right) \frac{1}{v} \left( 1+2 \frac{\Delta_a}{v} \right)
\end{equation}
and similarly for the term with $|{\bf y}-{\bf u}-{\bf a}|^3$ in (\ref{A1}). Expanding to first order in $a$ and keeping the linear terms we obtain
\begin{align}
	G^A_{1,i=3}=\frac{1}{v^2} \left( v_3 \cos\Omega_0 K \Delta_a+2\frac{v_3}{v} \sin\Omega_0 \Delta_a - a \sin\Omega_0 \right) \\
-\frac{1}{v^3} \left( -v_3 \sin\Omega_0 K \Delta_a+3\frac{v_3}{v} \cos\Omega_0 \Delta_a - a \cos\Omega_0 \right)
\end{align}

Then we proceed as in (\ref{ex27}), (\ref{ex28}) to obtain
\begin{equation}
E^A_i=-2qK^2 \int_M d^3u \int d^3y H^A({\bf x},{\bf y},{\bf u})\cdot aG^A_1(t,{\bf x},{\bf y},{\bf u}) + o(a^2)
\label{ex28perEA}
\end{equation}
The integrals (\ref{ex28}), (\ref{ex28perEA}) are performed via a standard Monte Carlo algorithm. 
Results (compared to a proper benchmark value, see Sect.\ \ref{bench}) are given in Sect.\ \ref{res}.
In the regions with $y_i<10^{-5}$ it is possible to compare numerically the integrals of some of the terms of the dipolar expansion with the corresponding terms of the full integrals (\ref{eq14}), (\ref{eq15}). This gives a cross-check of the dipolar expansion. Terms beyond the first order in $a$ are certainly not needed in our case, because the only significant contributions to the integrals come from the regions with $y_i>0.1$ cm (see Tab.\ \ref{tab2}), where the ratio $a/|{\bf y}|$ is very small.

\section{Benchmark values of $E_T$, $E_L$ from a conserved source}
\label{bench}

The numerical solution of the extended Maxwell equations found through the double-retarded integrals described in the previous Section, and whose raw results (only for $E_L$) are given in the Appendix, gives the components of the electric field in CGS units, referred to a source equal to 1 in the same units. From this solution we can see that $E_L \gg E_T$, and this certainly signals that something interesting occurs, compared to the usual propagation of $E_T$ only which occurs in the Maxwell theory with locally conserved sources. The absolute value of the fields, however, is little informative in itself and we need some benchmark. For this purpose we shall now compute the field generated, at the same position ($r=10$ cm, $\theta=45^{\circ}$) by a standard oscillating dipole with the same frequency and amplitude. By standard we mean that its current is locally conserved. A textbook formula for this case is
\begin{equation}
E_T=\frac{q \dot{v}\sin\theta}{c^2r}
\label{book}
\end{equation}
and yields an amplitude $E_T\simeq q \cdot 0.8 \cdot 10^{-7}$ (CGS units), supposing an harmonic oscillation with amplitude $a=2.5 \cdot 10^{-7}$ cm, $\omega=2\pi \cdot 10^{10}$ Hz. Since $E_L$ is of the order of $q \cdot 10^{-4}$ (see raw data in Tab.\ \ref{tab2} of the Appendix), this shows that the anomalous longitudinal field $E_L$ of an oscillating dipole with ``full'' strong tunnelling (i.e., one in which {\em all} charge oscillates between $-{\bf a}$ and ${\bf a}$ without an intermediate current) is about 2 or 3 orders of magnitude larger than the regular transverse field $E_T$ of a corresponding conserved source. 

In order to obtain a more precise estimate of the benchmark transverse field, we shall next compute it from the standard solution of the Maxwell equations with a source which is exactly equal to the source (\ref{dd-source}) ``completed'' with a current which ensures local conservation. This also makes the entire computation self-contained and yields a consistency check for the formalism employed. 

After writing the time derivative of the charge density $\rho$ in (\ref{dd-source}), we set it equal by definition to $-\nabla {\bf J}^c$ and obtain in this way the conserved current density ${\bf J}^c$.
It is straightforward to check that from the condition
\begin{align}
\frac{\partial \rho^c}{\partial t} =
	\frac{\partial}{\partial t} q \cos(\omega t) \left[ \delta^3(\textbf{x}-\textbf{a})-\delta^3(\textbf{x}+\textbf{a}) \right] \equiv - \frac{\partial}{\partial x_3} J^c_3
\end{align}
one has
\begin{align}
	J_3^c=-q \omega\sin(\omega t) \left[\theta(x_3+a)+\theta(-x_3+a)-1 \right] \delta(x_1) \delta(x_2)
\end{align}

The standard Maxwell equations in Lorenz gauge in CGS units for the potentials $\phi^c$, $\textbf{A}^c$ are (the subscript \emph{c} stays for ``conserved'')
\begin{align}
	\frac{1}{c^2} \frac{\partial^2 \phi^c}{\partial t^2} - \nabla^2 \phi^c=4\pi\rho^c
\end{align}
\begin{align}
	\frac{1}{c^2} \frac{\partial^2 \textbf{A}^c}{\partial t^2} - \nabla^2 \textbf{A}^c=\frac{4\pi}{c}\textbf{J}^c
\end{align}
and their solutions ($k=c^{-1}$)
\begin{align}
	\phi^c(\textbf{x},t)=\int d^3y \frac{1}{|\textbf{x}-\textbf{y}|}\rho^c(\textbf{y},t-k|\textbf{x}-\textbf{y}|) \\
	\textbf{A}^c(\textbf{x},t)=\int d^3y \frac{1}{|\textbf{x}-\textbf{y}|}\textbf{J}^c(\textbf{y},t-k|\textbf{x}-\textbf{y}|)
\end{align}

The integral for $\phi^c$ gives
\begin{align}
	\phi^c=\frac{q}{|\textbf{x}-\textbf{a}|} \cos\left[ \omega (t-k|\textbf{x}-\textbf{a}|) \right]-
	\frac{q}{|\textbf{x}+\textbf{a}|} \cos\left[ \omega (t-k|\textbf{x}+\textbf{a}|) \right]
\label{eq56}
\end{align}
The corresponding contribution to the electric field is obtained from $-\nabla\phi$. 

The integral for $A^c_3$ gives (the other components of $\textbf{A}^c$ vanish)
\begin{align}
	A^c_3(\textbf{x},t)=-qk \int dy_3 \left[ \frac{\omega\sin\left[\omega(t-k|\textbf{x}-\textbf{y}|)\right]}{|\textbf{x}-\textbf{y}|} \right]_{y_1=y_2=0} \left[\theta(y_3+a)+\theta(-y_3+a)-1\right]
\end{align}
The corresponding contribution to the electric field is obtained with $-k\partial_t$ and is
\begin{align}
	E^{A,c}_3(\textbf{x},t)=q(k\omega)^2 \int_{-a}^a ds \frac{\cos\left[\omega(t-k\sqrt{x_1^2+x_2^2+(x_3-s)^2})\right]}{\sqrt{x_1^2+x_2^2+(x_3-s)^2}}
\end{align}

These formulas allow to obtain the components $E^c_T$, $E^c_L$, taking into account that we have fixed for simplicity $\theta=45^{\circ}$. Setting the distance at $r=10$ cm for comparison with the anomalous fields, we can compute the field components for different values of $t$. Since all components oscillate at high frequency, we take the root mean square of $E^c_T$, $E^c_L$ over many values of $t$. With 1000 values we obtain 
\begin{align}
	q^{-1}E_T^{c,rms}=1.54\cdot 10^{-7}; \qquad \qquad 
	q^{-1}E_L^{c,rms}=1.49\cdot 10^{-8} \ \ \ {\rm (CGS \ units)}
\end{align}

As expected, $E^c_L \ll E^c_T$, since we are at a distance $r\simeq 3\lambda$.

\section{Results of the retarded integrals for the anomalous longitudinal field. Discussion}
\label{res}

\subsection{Dependence on the distance}

\begin{table}[h]
\begin{center}
\begin{tabular}{|c|c|c|c|}
\hline
\ & $r=10$ cm & $r=10.75$ cm & $r=11.5$ cm  \\
\ & ($\sim 3\lambda$) & ($10$ cm + $\lambda/4$) & ($10$ cm + $\lambda/2$)  \\
\hline
$\frac{E_L(r)}{E_T^{c,rms}(10)}$ & $9.5 \cdot 10^{2}$ & $-6.5 \cdot 10^{2}$ & $-10.8 \cdot 10^{2}$ \\
\hline
\end{tabular}
\end{center}
\caption
{Normalized longitudinal field at $t=0$ as a function of the distance $r$. See explanations in the main text. Source parameters: $a=2.5\cdot 10^{-7}$ cm, $d=\varepsilon=10^{-7}$ cm, $f=10^{10}$ Hz.}
\label{tab1}
\end{table}

The double-retarded integrals, computed numerically as described in the previous sections, give a longitudinal (i.e., radial) component of the electric field of the order of $10^2$ to $10^3$ times greater than the standard transverse component. The computation is done along a radial line forming an angle $\theta=45^{\circ}$ with respect to the oscillation axis of the dipole. Initially we take the oscillation frequency equal to $10^{10}$ Hz and the field is computed at the instant $t=0$. In order to check that the two field components $E_T$, $E_L$ oscillate in the wave zone with a wavelength $\lambda=3$ cm, corresponding to the chosen frequency, we have computed these components at distance $r=10$ cm and then increased the distance in steps of $\lambda/4$. The results (Tab.\ \ref{tab1}) actually show an oscillating behavior as expected, within the uncertainties. All values in the table are normalized to $E_T^{c,rms}(10)$, which is the r.m.s.\ value of the transverse field generated at distance $r=10$ cm by a standard ``completed'' oscillating dipole, as described in Sect.\ \ref{bench}. The values of $E_T$ obtained for the anomalous source are of the same magnitude order as those of the standard source, and are not reported in the table. 

With a further increase in the distance we then pass to $r=25$ cm (10 cm + 5$\lambda$) and $r=40$ cm (10 cm + 10$\lambda$). Like for $r=10$ cm, we find values of $E_L$ close to the maxima of the oscillation, but the oscillation amplitude appears to have increased: we have respectively 
\begin{align}
	\frac{E_L(25)}{E_T^{c,rms}(10)}\simeq 2.2\cdot 10^3; \qquad \qquad
	\frac{E_L(40)}{E_T^{c,rms}(10)}\simeq 3.4\cdot 10^3
\end{align}

The raw data (non normalized) of the contributions to $E_L$ coming from the various integration regions (Tab.\ \ref{tab2} of the Appendix) show that the secondary charge which generates the large values of $E_L$ at $r=25$ cm and $r=40$ cm is located farther away from the dipole, in comparison to the secondary charge generating $E_L$ at $r=10$ cm. In other words, as we move farther away from the dipole the longitudinal field increases because it is generated by a larger portion of the ``cloud'' of secondary charge. It is not easy, however, to understand intuitively exactly how  the different portions of the cloud contribute to the field, because we are not in a stationary state, but everything oscillates at high frequency, including the charge cloud itself, and the phase $\Omega=\omega(t-k|{\bf x}-{\bf y}|-k|{\bf y}-{\bf z}|)$ in the integrals (\ref{eq14}), (\ref{eq15}) produces double-retarded interference effects.

From the present data it is not possible to assess the behavior of the longitudinal field at greater distances, because the uncertainties in the integrals in the regions with distance above approx.\ 100 cm are too large. We expect, of course, an eventual decrease of $E_L$. 

Notice that while the $E_T$ component of a standard e.m.\ radiation field must decrease steadily ar $1/r$ in order to maintain the Poynting flux constant, the $E_L$ component does not contribute to this flux. However, it is not clear yet, in our opinion, what are the correct expressions for the e.m.\ densities of energy and momentum in the extended Aharonov-Bohm electrodynamics, even though this issue has been addressed in some of the cited works.

\subsection{Dependence on time and on the shape of the sources}

Concerning the dependence on time, at fixed distance, we have checked that it is periodic as expected, with frequency $\omega$. For instance, at the time $t=0.5\cdot 10^{-10}$ s, the figures of Tab.\ \ref{tab1} change signs.

In order to vary the shape of the sources, we change the parameter $d$ in the Gaussian charge density Ansatz (\ref{doppia-exp}); this parameter fixes the size of the source in the directions $z_1$ and $z_2$, i.e.\ transversally with respect to the oscillation direction of the dipole. The data in Tab.\ \ref{tab1} have been obtained setting $d=\varepsilon=10^{-7}$ cm, thus with sources having spherical symmetry.

One observes that the longitudinal emission is independent from $d$, at least up to $d=20\cdot 10^{-7}$ cm, which corresponds in practice to having two parallel discs instead of two pointlike sources. For practical applications in superconductors this is important, because wide junctions are more likely to carry a large current, in comparison to pointlike contacts. This independence from $d$ at high frequency should be contrasted with the behavior of the anomalous magnetic field in the quasi-static case \cite{modanese2019design}: in that case, increasing $d$ rapidly leads to the suppression of the anomaly.

\subsection{Dependence on the frequency}

The choice of the oscillation frequency in the calculation is crucial because it defines the wavelength and therefore the integration regions. The value $f=10^{10}$ Hz seems to be a good compromise, because such a frequency can be easily obtained in the self-oscillation of a Josephson junction and is still accessible as an external bias for a molecular nano-device.

We also made some variations of $f$ in the calculation. Setting for instance $f=0.5\cdot 10^{10}$ Hz, we evaluated the longitudinal field at distance $r=20$ cm, which corresponds to little more than $3\lambda$ (like $r=10$ cm for $f=10^{10}$ Hz), and similarly with $f=2\cdot 10^{10}$ Hz. In each case, the value of $E_L$ found was compared to the r.m.s.\ of $E^c_T$ at the same distance for a standard conserved source. The resulting ratios show only a weak dependence on the frequency in this range.

\section{Conclusions}
\label{conc}

At the level of fundamental interactions there are no doubts on the full validity of quantum field theory, and in particular of QED and of the principle of local charge conservation. Nevertheless, in the presence of non-local interactions (either as an effective descriptive model, or with fundamental motivations like in fractional quantum mechanics), the failure of local conservation of the ``$\rho v$ current'' inevitably leads to a new ``emergent'' phenomenology, characterized by secondary currents which may extend outside the primary source and generate non-standard fields. The real physical properties of these secondary currents are not yet properly understood. We think that experiments will play a fundamental role in clarifying this issue. In our latest work \cite{modanese2019design} we proposed a design of a device for the detection of anomalous magnetic fields generated by quasi-stationary non-conserved currents. For the case of an high-frequency oscillating source considered in this paper the choice of the experimental strategy is more obvious, namely a search for longitudinal electric fields in the radiation zone. We plan to discuss this in more details in forthcoming work. 

Another crucial question is, for which materials the non-local part of the current is expected to achieve the level sufficient for detection (at least 1 part in $10^5$, if we admit for instance that a longitudinal field of the order of 1\% of the transverse field can be safely detected).

The choice of the dipole length $a$ for our numerical solution has been motivated by a possible application to Josephson tunnelling in YBCO. In the case of molecular nano-devices the typical sizes and shapes of current sources and sinks arising in the case of local non-conservation should be estimated through the density functional theory; on the experimental side, trials with, e.g., graphene antennas emitting in the GHz range could give useful insights. 

\bigskip

\noindent
{\bf Acknowledgment} - This work was supported by the Open Access Publishing Fund of the Free University of Bozen-Bolzano.

\section{Appendix}

\subsection{Raw results of the Monte Carlo integration for the longitudinal field component}

\begin{table}[h]
\begin{center}
\begin{tabular}{|c|c|c|c|c|c|}
\hline
\textbf{Integr.\ region} & $r=10$ cm & $r=10.75$ & $r=11.5$ & $r=25$ & $r=40$ \\
\hline
$[0.1,1]$ cm & 6 & 1 & -5 & 2 & 1  \\
$[1,10]$ & 144 & -98 & -159 & 107 & 60  \\
$[10,40]$ & -2 & -2 & -3 & 225 & 455 \\
$[40,100]$ & $0\pm 1$ & $0\pm 5$ & $0\pm 3$ & $0\pm 4$ & $0\pm 10$  \\
\hline
\end{tabular}
\end{center}
\caption
{Contributions to the longitudinal electric field from the four main integration regions of the ${\bf y}$ variable in the integrals (\ref{ex28}), (\ref{ex28perEA}). Data in CGS units, multiplied by $10^6$ and referred to a source charge $q=1$. For the total field properly normalized to a transverse component see Sect.\ \ref{res}. The integration regions with $y_i<0.1$ cm and $y_i>100$ cm do not give significant contributions. The values of the transverse field are not reported and are typically of the order of 1, in the same units, or less. The C code used is appended at the end of the TeX source of this paper.}
\label{tab2}
\end{table}

\subsection{Aharonov-Bohm-Maxwell extended equations in CGS units}

The equations without sources are written as usual, namely $\nabla \times \textbf{E}=-(1/c)(\partial \textbf{B}/\partial t)$, $\nabla \cdot \textbf{B}=0$. The extended equations with sources take the form
\begin{align}
	\nabla \cdot \textbf{E}=4\pi \rho-\frac{1}{c^2}\frac{\partial}{\partial t}\int d^3y \frac{I\left(t_{ret},\textbf{y} \right)}{\left|\textbf{x}-\textbf{y} \right|} \label{eqE}
\end{align}
\begin{align}
	\nabla \times \textbf{B}-\frac{1}{c} \frac{\partial \textbf{E}}{\partial t}=\frac{4\pi}{c} \textbf{J}+\frac{1}{c} \nabla \int d^3y \frac{I\left(t_{ret},\textbf{y} \right)}{\left|\textbf{x}-\textbf{y} \right|} \label{eqB}
\end{align}
where $I_{ret}=I(t-|\textbf{x}-\textbf{y}|/c,\textbf{x})$ and $I=\partial_t \rho+\nabla\cdot\textbf{J}$.

\bibliographystyle{unsrt}
\bibliography{mme}

\begin{thebibliography}{10}

\bibitem{ohmura1956new}
T.~Ohmura.
\newblock A new formulation on the electromagnetic field.
\newblock {\em Progress of Theoretical Physics}, 16(6):684--685, 1956.

\bibitem{aharonov1963further}
Y.~Aharonov and D.~Bohm.
\newblock Further discussion of the role of electromagnetic potentials in the
  quantum theory.
\newblock {\em Physical Review}, 130(4):1625, 1963.

\bibitem{van2001generalisation}
K.J. Van~Vlaenderen and A.~Waser.
\newblock Generalisation of classical electrodynamics to admit a scalar field
  and longitudinal waves.
\newblock {\em Hadronic Journal}, 24(5):609--628, 2001.

\bibitem{jimenez2011cosmological}
J.B. Jim{\'e}nez and A.L. Maroto.
\newblock Cosmological magnetic fields from inflation in extended
  electromagnetism.
\newblock {\em Physical Review D}, 83(2):023514, 2011.

\bibitem{hively2012toward}
L.M. Hively and G.C. Giakos.
\newblock Toward a more complete electrodynamic theory.
\newblock {\em International Journal of Signal and Imaging Systems
  Engineering}, 5(1):3--10, 2012.

\bibitem{Modanese2017MPLB}
G.~Modanese.
\newblock {Generalized Maxwell equations and charge conservation censorship}.
\newblock {\em Modern Physics Letters B}, 31:1750052, 2017.

\bibitem{modanese2017electromagnetic}
G.~Modanese.
\newblock Electromagnetic coupling of strongly non-local quantum mechanics.
\newblock {\em Physica B: Condensed Matter}, 524:81--84, 2017.

\bibitem{arbab2017extended}
A.I. Arbab.
\newblock Extended electrodynamics and its consequences.
\newblock {\em Modern Physics Letters B}, 31(09):1750099, 2017.

\bibitem{modanese2019design}
G.~Modanese.
\newblock Design of a test for the electromagnetic coupling of non-local
  wavefunctions.
\newblock {\em Results in Physics}, 12:1056--1061, 2019.

\bibitem{hively2019classical}
L.M. Hively and A.S. Loebl.
\newblock Classical and extended electrodynamics.
\newblock {\em Physics Essays}, 32(1):112--126, 2019.

\bibitem{laskin2002fractional}
N.~Laskin.
\newblock {Fractional Schr{\"o}dinger equation}.
\newblock {\em Physical Review E}, 66(5):056108, 2002.

\bibitem{Lenzi2008fractional}
E.K. Lenzi, B.F. De~Oliveira, N.G.C. Astrath, L.C. Malacarne, R.S. Mendes, M.L.
  Baesso, and L.R. Evangelista.
\newblock Fractional approach, quantum statistics, and non-crystalline solids
  at very low temperatures.
\newblock {\em The European Physical Journal B-Condensed Matter and Complex
  Systems}, 62(2):155--158, 2008.

\bibitem{zhang2015propagation}
Y.~Zhang, X.~Liu, M.R. Beli{\'c}, W.~Zhong, Y.~Zhang, M.~Xiao, et~al.
\newblock {Propagation dynamics of a light beam in a fractional Schr{\"o}dinger
  equation}.
\newblock {\em Physical Review Letters}, 115(18):180403, 2015.

\bibitem{wei2016comment}
Y.~Wei.
\newblock {Comment on ``Fractional quantum mechanics'' and ``Fractional
  Schr{\"o}dinger equation''}.
\newblock {\em Physical Review E}, 93(6):066103, 2016.

\bibitem{zhang2017unveiling}
D.~Zhang, Y.~Zhang, Z.~Zhang, N.~Ahmed, Y.~Zhang, F.~Li, M.R. Beli{\'c}, and
  M.~Xiao.
\newblock {Unveiling the link between fractional Schr{\"o}dinger equation and
  light propagation in honeycomb lattice}.
\newblock {\em Annalen der Physik}, 529(9):1700149, 2017.

\bibitem{petreska2019time}
I.~Petreska, A.S.M. de~Castro, T.~Sandev, and E.K. Lenzi.
\newblock {The time-dependent Schr{\"o}dinger equation in three dimensions
  under geometric constraints}.
\newblock {\em Journal of Mathematical Physics}, 60(3):032101, 2019.

\bibitem{modanese2018time}
G.~Modanese.
\newblock Time in quantum mechanics and the local non-conservation of the
  probability current.
\newblock {\em Mathematics}, 6(9):155, 2018.

\bibitem{baraff1998model}
G.A. Baraff.
\newblock Model for the effect of finite phase-coherence length on resonant
  transmission and capture by quantum wells.
\newblock {\em Physical Review B}, 58(20):13799, 1998.

\bibitem{ferry1999complex}
D.K. Ferry, J.R. Barker, and R.~Akis.
\newblock Complex potentials, dissipative processes, and general quantum
  transport.
\newblock In {\em Proceedings of the 1999 International Conference on Modelling
  and Simulation of Micro Systems, NSTI}, pages 373--376, 1999.

\bibitem{chamon1997nonlocal}
L.C. Chamon, D.~Pereira, M.S. Hussein, M.A.C. Ribeiro, and D.~Galetti.
\newblock Nonlocal description of the nucleus-nucleus interaction.
\newblock {\em Physical Review Letters}, 79(26):5218, 1997.

\bibitem{balantekin1998green}
A.B. Balantekin, J.F. Beacom, et~al.
\newblock Green's function for nonlocal potentials.
\newblock {\em Journal of Physics G: Nuclear and Particle Physics},
  24(11):2087, 1998.

\bibitem{latora1999superdiffusion}
V.~Latora, A.~Rapisarda, and S.~Ruffo.
\newblock Superdiffusion and out-of-equilibrium chaotic dynamics with many
  degrees of freedoms.
\newblock {\em Physical Review Letters}, 83(11):2104, 1999.

\bibitem{caspi2000enhanced}
A.~Caspi, R.~Granek, and M.~Elbaum.
\newblock Enhanced diffusion in active intracellular transport.
\newblock {\em Physical Review Letters}, 85(26):5655, 2000.

\bibitem{Lenzi2008solutions}
E.K. Lenzi, B.F. de~Oliveira, L.R. da~Silva, and L.R. Evangelista.
\newblock {Solutions for a Schr{\"o}dinger equation with a nonlocal term}.
\newblock {\em Journal of Mathematical Physics}, 49(3):032108, 2008.

\bibitem{sandev2014time}
T.~Sandev, I.~Petreska, and E.K. Lenzi.
\newblock {Time-dependent Schr{\"o}dinger-like equation with nonlocal term}.
\newblock {\em Journal of Mathematical Physics}, 55(9):092105, 2014.

\bibitem{sandev2016effective}
T.~Sandev, I.~Petreska, and E.K. Lenzi.
\newblock {Effective potential from the generalized time-dependent
  Schr{\"o}dinger equation}.
\newblock {\em Mathematics}, 4(4):59, 2016.

\bibitem{li2008definition}
C.~Li, L.~Wan, Y.~Wei, and J.~Wang.
\newblock Definition of current density in the presence of a non-local
  potential.
\newblock {\em Nanotechnology}, 19(15):155401, 2008.

\bibitem{zhang2011first}
L.~Zhang, B.~Wang, and J.~Wang.
\newblock First-principles calculation of current density in molecular devices.
\newblock {\em Physical Review B}, 84(11):115412, 2011.

\bibitem{dreyer2018current}
C.E. Dreyer, M.~Stengel, and D.~Vanderbilt.
\newblock Current-density implementation for calculating flexoelectric
  coefficients.
\newblock {\em Physical Review B}, 98(7):075153, 2018.

\bibitem{hook1973ginzburg}
J.R. Hook and J.R. Waldram.
\newblock {A Ginzburg-Landau equation with non-local correction for
  superconductors in zero magnetic field}.
\newblock {\em Proc. R. Soc. Lond. A}, 334(1597):171--192, 1973.

\bibitem{hilgenkamp2002grain}
H.~Hilgenkamp and J.~Mannhart.
\newblock {Grain boundaries in high-Tc superconductors}.
\newblock {\em Reviews of Modern Physics}, 74(2):485, 2002.

\bibitem{botello2011quantum}
A.R. Botello-M{\'e}ndez, E.~Cruz-Silva, J.M. Romo-Herrera,
  F.~L{\'o}pez-Ur{\'\i}as, M.~Terrones, B.G. Sumpter, H.~Terrones, J.-C.
  Charlier, and V.~Meunier.
\newblock Quantum transport in graphene nanonetworks.
\newblock {\em Nano Letters}, 11(8):3058--3064, 2011.

\bibitem{giakos1993detection}
G.C. Giakos and T.~Ishii.
\newblock Detection of longitudinal electromagnetic fields in air.
\newblock {\em Microwave and Optical Technology Letters}, 6(5):283--287, 1993.

\bibitem{monstein2002observation}
C.~Monstein and J.-P. Wesley.
\newblock Observation of scalar longitudinal electrodynamic waves.
\newblock {\em EPL (Europhysics Letters)}, 59(4):514, 2002.

\bibitem{monstein2004remarks}
C.~Monstein and J.P. Wesley.
\newblock {Remarks to the Comment by J.R.\ Bray and M.C.\ Britton on
  ``Observation of scalar longitudinal electrodynamic waves''}.
\newblock {\em EPL (Europhysics Letters)}, 66(1):155, 2004.

\end{thebibliography}

\end{document}